\documentclass[conference]{IEEEtran}
\usepackage{cite}
\ifCLASSINFOpdf
   \usepackage[pdftex]{graphicx}
   \DeclareGraphicsExtensions{.pdf,.jpeg,.png}
\else
   \usepackage[dvips]{graphicx}
   \DeclareGraphicsExtensions{.eps}
\fi
\usepackage[cmex10]{amsmath}
\usepackage{algorithmic}
\usepackage{array}
\usepackage{url}
\usepackage{amsthm}
\theoremstyle{plain}\newtheorem{theo}{Theorem}

\begin{document}
%
% paper title
% can use linebreaks \\ within to get better formatting as desired
\title{Discussion of Twenty Questions Problem}

% author names and affiliations
% use a multiple column layout for up to three different
% affiliations
\author{\IEEEauthorblockN{Barco You}
\IEEEauthorblockA{Department of Electronics and Information Engineering\\
Huazhong University of Science and Technology\\
Wuhan, China 430074\\
Email: barcojie@gmail.com}
}
%\and
%\IEEEauthorblockN{Homer Simpson}
%\IEEEauthorblockA{Twentieth Century Fox\\
%Springfield, USA\\
%Email: homer@thesimpsons.com}
%\and
%\IEEEauthorblockN{James Kirk\\ and Montgomery Scott}
%\IEEEauthorblockA{Starfleet Academy\\
%San Francisco, California 96678-2391\\
%Telephone: (800) 555--1212\\
%Fax: (888) 555--1212}}

% conference papers do not typically use \thanks and this command
% is locked out in conference mode. If really needed, such as for
% the acknowledgment of grants, issue a \IEEEoverridecommandlockouts
% after \documentclass

% for over three affiliations, or if they all won't fit within the width
% of the page, use this alternative format:
% 
%\author{\IEEEauthorblockN{Michael Shell\IEEEauthorrefmark{1},
%Homer Simpson\IEEEauthorrefmark{2},
%James Kirk\IEEEauthorrefmark{3}, 
%Montgomery Scott\IEEEauthorrefmark{3} and
%Eldon Tyrell\IEEEauthorrefmark{4}}
%\IEEEauthorblockA{\IEEEauthorrefmark{1}School of Electrical and Computer Engineering\\
%Georgia Institute of Technology,
%Atlanta, Georgia 30332--0250\\ Email: see http://www.michaelshell.org/contact.html}
%\IEEEauthorblockA{\IEEEauthorrefmark{2}Twentieth Century Fox, Springfield, USA\\
%Email: homer@thesimpsons.com}
%\IEEEauthorblockA{\IEEEauthorrefmark{3}Starfleet Academy, San Francisco, California 96678-2391\\
%Telephone: (800) 555--1212, Fax: (888) 555--1212}
%\IEEEauthorblockA{\IEEEauthorrefmark{4}Tyrell Inc., 123 Replicant Street, Los Angeles, California 90210--4321}}

% use for special paper notices
%\IEEEspecialpapernotice{(Invited Paper)}

% make the title area
\maketitle

\begin{abstract}
%\boldmath
Discuss several tricks for solving twenty question problems which in this
paper is depicted as a guessing game. Player tries to find a ball in twenty 
boxes by asking as few questions as possible, and these questions are 
answered by only ``Yes'' or ``No''. With the discussion, demonstration of source coding methods is the main concern.
\end{abstract}
% IEEEtran.cls defaults to using nonbold math in the Abstract.
% This preserves the distinction between vectors and scalars. However,
% if the conference you are submitting to favors bold math in the abstract,
% then you can use LaTeX's standard command \boldmath at the very start
% of the abstract to achieve this. Many IEEE journals/conferences frown on
% math in the abstract anyway.

% no keywords

% For peer review papers, you can put extra information on the cover
% page as needed:
% \ifCLASSOPTIONpeerreview
% \begin{center} \bfseries EDICS Category: 3-BBND \end{center}
% \fi
%
% For peerreview papers, this IEEEtran command inserts a page break and
% creates the second title. It will be ignored for other modes.
\IEEEpeerreviewmaketitle

\section{Introduction}
% no \IEEEPARstart
\indent Unit computation of mordern computer is still binary, while ``Yes or No''
question is a good illustration of such computing, asking one question is equivalent
to spending one bit of computation resource. This discussion 
is intended to give an intution behind symbol source coding through discussing 
the different ways for solving a concrete twenty question problem.\\ 
\indent The rest of this paper is organized as follows. Section \ref{sec:2}{} introduces the way of
one-by-one asking. Section \ref{sec:3}{} is about top-down division. In Section \ref{sec:4}{}
we discuss the way of down-top merging. The work is concluded in Section \ref{sec:5}.

\section{One-bye-One Asking}\label{sec:2}
\begin{figure}[!h]
\centering
\includegraphics[width=2.5in]{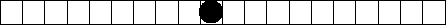}
\caption{Only one of twenty boxes includes a ball}
\label{fig:1}
\end{figure}

\begin{figure}[t]
  \centering
  \includegraphics[width=2.5in]{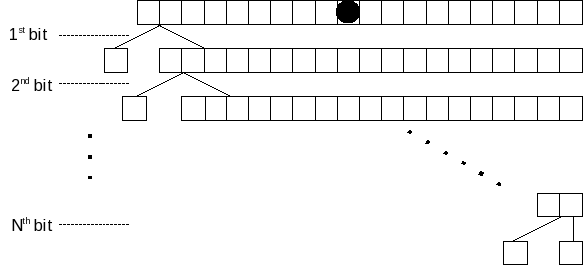}
  \caption{Illustration of One-by-One Asking}
  \label{fig:2}
\end{figure}

\indent We depict the TQP(Twenty Question Problem) with 20 boxes in which only one box contains a ball,
shown as figure~\ref{fig:1}. With method one, we choose arbitraily one box and say
it contain the ball, if opening the box and find there is none, equivalently answered
by ``No'', we get information content $\log{\frac{20}{19}}$. Continuously we draw 
another box but miss the ball again, we get information content $\log{\frac{19}{18}}$.
Step forward repeatedly, and assume the ball is found at step $N (1\leq N \leq 20)$, up to now the total
information content we got is $(\log{\frac{20}{19}}+\log{\frac{19}{18}}+\cdots+%
\log{\frac{20-N+2}{20-N+1}}+\log{\frac{20-N+1}{1}} = \log{\frac{20}{1}} = 4.3219 bits)$.
\\
\indent Without loss of generality, the guessing process is illustrated as choosing the
boxes in order from left to right, shown as figure~\ref{fig:2}. For every guessing, we
have ``Yes'' or ``No'' results, Imagine that 1 bit is spent for every guessing. Then
the expected bits need solving the TQP with the One-by-One method equals to $(1+\frac{19}{20}+%
\frac{18}{20}+\cdots+\frac{2}{20} = \frac{209}{20} = 10.45 bits)$.
\section{Top-Down Division}\label{sec:3}
\begin{figure}[t]
  \centering
  \includegraphics[width=2.5in]{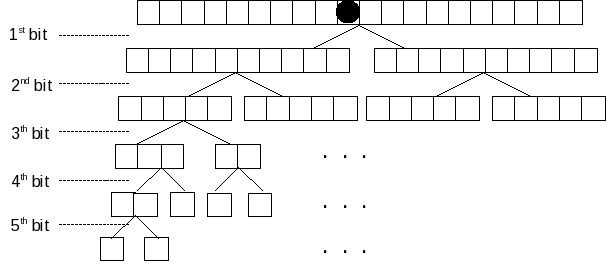}
  \caption{Illustration of Top-Down Division}
  \label{fig:3}
\end{figure}

\indent Before every asking we divede equally the boxes into two groups, then ask if the ball is in
one of the two groups. According to the answer continue this strategy repeatly until the ball is found.
This division process is shown as figure~\ref{fig:3}. In this way the expected bits to spend is
$(1+1+1+1+1\times\frac{8}{20} = 4.4 bits)$.\\
\indent The information content gotten from this ways is $(1+(\frac{10}{20}+\frac{10}{20})+\frac{5}{20}\times%
(\frac{3}{5}\log{\frac{5}{3}}+\frac{2}{5}\log{\frac{5}{2}})\times4+\frac{2}{20}\times4+\frac{3}{20}\times%
(\frac{2}{3}\log{\frac{3}{2}}+\frac{1}{3}\log{3})\times4+\frac{2}{20}\times4 = \log{20} = 4.3219 bits)$.
\section{Down-Top Merging}\label{sec:4}
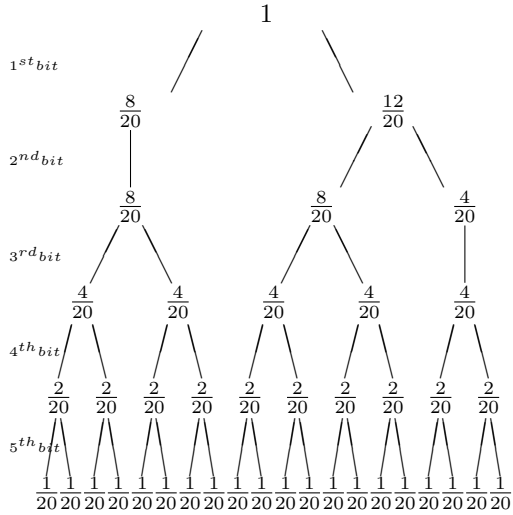
\begin{figure}[t]
  \centering
  \setlength{\unitlength}{0.1in}
  \begin{picture}(25,25)
	\multiput(1.25,0)(1.25,0){20}{$\frac{1}{20}$}
	\multiput(1.875,5)(2.5,0){10}{$\frac{2}{20}$}
	\multiput(3.125,10)(5,0){5}{$\frac{4}{20}$}
	\multiput(5.625,15)(10,0){2}{$\frac{8}{20}$}
	\put(23.125,15){$\frac{4}{20}$}
	\put(5.625,20){$\frac{8}{20}$}
	\put(19.375,20){$\frac{12}{20}$}
	\put(13.125,25){$1$}
	\multiput(1.95,1.25)(2.5,0){10}{\line(1,6){0.5}}
	\multiput(3.2,1.25)(2.5,0){10}{\line(-1,6){0.5}}
	\multiput(2.575,6.35)(5,0){5}{\line(1,4){0.7}}
	\multiput(5.075,6.35)(5,0){5}{\line(-1,4){0.7}}
	\multiput(4.25,11.35)(10,0){2}{\line(1,2){1.5}}
	\multiput(8.5,11.35)(10,0){2}{\line(-1,2){1.5}}
	\put(23.85,11.35){\line(0,1){3}}
	\put(6.35,16.35){\line(0,1){3}}
	\put(17.35,16.35){\line(1,2){1.6}}
	\put(22.75,16.35){\line(-1,2){1.6}}
	\put(8.5,21.35){\line(1,2){1.6}}
	\put(18,21.35){\line(-1,2){1.6}}
	\put(0,2.5){\tiny{$5^{th}bit$}}
	\put(0,7.5){\tiny{$4^{th}bit$}}
	\put(0,12.5){\tiny{$3^{rd}bit$}}
	\put(0,17.5){\tiny{$2^{nd}bit$}}
	\put(0,22.5){\tiny{$1^{st}bit$}}
  \end{picture}
  \caption{Illustration of Down-Top Merging}
  \label{fig:4}
\end{figure}

\indent The smartest way presented here is to merge the options in Down-Top direction, which follows
Huffman Coding method\cite{ref:1}. Every box has the same probability $\frac{1}{20}$ to contain the ball, combine
two of the boxes and imagine they become a bigger one, then the probability of the ball in this bigger box is
$\frac{2}{20}$. For every merging we make sure that the two boxes (real or imagined box) have the smallest
probability of including the ball. For example, after first merging we have one bigger box which has probability
$\frac{2}{20}$ and there are 18 boxes with probability $\frac{1}{20}$, so 9 bigger boxes should be formed from
the 18 boxes respectively. Repeat merging bigger boxes until we have a box which include the ball with probability
$1$. This merging process is shown as figure~\ref{fig:4}. From this process we have the spent bits is $(1+\frac{12}%
{20}+(\frac{8}{20}\times 2)+(\frac{4}{20}\times 5)+(\frac{2}{20}\times 10) = 4.4 bits)$.\\
\indent The information content gotten in this way is $(1\times(\frac{8}{20}\log{\frac{20}{8}}+\frac{12}{20}%
\log{\frac{20}{12}})+\frac{12}{20}\times(\frac{8}{12}\log{\frac{12}{8}}+%
\frac{4}{12}\log{\frac{12}{4}})+\frac{8}{20}\times1\times2+\frac{4}{20}\times1\times5+\frac{2}{20}\times1\times10%
= 4.3219 bits)$.
\section{Conclusion}\label{sec:5}
\indent From above discussion, we can definitely conclude that to find the ball the three tricks get the same information
content, but the first method consume in average much more extra effort than the later two methods. 
For TQP, the Top-Down Divsion method and 
Down-Top Merging method consume the same expected bits for achieving the goal. But they are not of the same efficiency.
Actually the Down-Top Merging is optimal while Top-Down Divsion is sub-optimal, just like nuclear fusion has much more
energy than nuclear fission.\\
\begin{theo}
  For symbol coding, Huffman code is the optimal.
\end{theo}
\begin{IEEEproof}
  Let symbol set $\mathcal{A}_X = \{x_1,\cdots,x_N\}$ have $\mathcal{P}_X = \{p_1,\cdots,p_N\}$. Use division or merging
  method to construct codes for symbols, with once division or merging we have a new level. At any level $I$ there are
  intermediate symbols $\mathcal{A}_I = \{\alpha_1,\cdots,\alpha_{n_I}\} (2\leq n_I \leq N)$,
  and $\mathcal{P}_I = \{p_1,\cdots,p_{n_I}\} (\sum_{k=1}^{n_I} p_k = 1)$. With Huffman coding method, at level $I$ we
  merge two symbols $\alpha_i \text{ and } \alpha_j$, $\forall k \in \{1,\cdots,n_i\}\text{ and }k\neq i, k\neq j:\qquad p_k 
  \geq p_i,p_j$. Then the bits consumed by this merge is $1\times(p_i+p_j)$. With other code, at any level $I$ if two
  symbols $\alpha_{k_1} \text{ and }\alpha_{k_2}$ merge into or are divided from $(I-1)$ level. The consumed bits 
  $1\times(p_{k_1}+p_{k_2})\geq1\times(p_i+p_j)$, if $k_1, k_2 \neq i, j$. Sum all the bits consumed at all levels, 
  we can get the Huffman code is the shortest.\\
\end{IEEEproof}

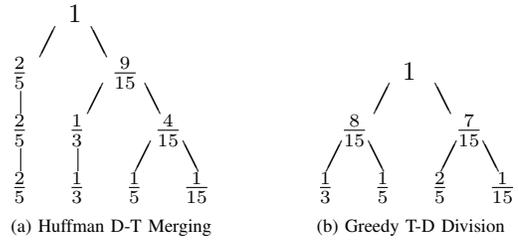
\begin{figure}[t]
\centering
\setlength{\unitlength}{0.1in}
\begin{picture}(25,12)
  \put(0,2){$\frac{2}{5}$}
  \put(3,2){$\frac{1}{3}$}
  \put(6,2){$\frac{1}{5}$}
  \put(9,2){$\frac{1}{15}$}
  \put(0,5){$\frac{2}{5}$}
  \put(3,5){$\frac{1}{3}$}
  \put(7.5,5){$\frac{4}{15}$}
  \put(0,8){$\frac{2}{5}$}
  \put(5.25,8){$\frac{9}{15}$}
  \put(3,11){$1$}
  \put(0.5,3.2){\line(0,1){1.2}}
  \put(3.5,3.2){\line(0,1){1.2}}
  \put(6.5,3.2){\line(1,2){0.8}}
  \put(9.8,3.2){\line(-1,2){0.8}}
  \put(0.5,6.2){\line(0,1){1.2}}
  \put(4,6.2){\line(1,2){0.8}}
  \put(7.8,6.2){\line(-1,2){0.8}}
  \put(1.5,9.2){\line(1,2){0.8}}
  \put(5,9.2){\line(-1,2){0.8}}
  \put(0,0){\scriptsize{(a) Huffman D-T Merging}}
  \put(16,2){$\frac{1}{3}$}
  \put(19,2){$\frac{1}{5}$}
  \put(22,2){$\frac{2}{5}$}
  \put(25,2){$\frac{1}{15}$}
  \put(17.25,5){$\frac{8}{15}$}
  \put(23.25,5){$\frac{7}{15}$}
  \put(20.5,8){$1$}
  \put(16.5,3.2){\line(1,2){0.8}}
  \put(19.5,3.2){\line(-1,2){0.8}}
  \put(22.5,3.2){\line(1,2){0.8}}
  \put(25.5,3.2){\line(-1,2){0.8}}
  \put(19,6.2){\line(1,2){0.8}}
  \put(23,6.2){\line(-1,2){0.8}}
  \put(16,0){\scriptsize{(b) Greedy T-D Division}}
\end{picture}
\caption{Comparison between Huffman method and Greedy division}
\label{fig:5}
\end{figure}

\indent Take an example as figure~\ref{fig:5}. A symbol set with $\mathcal{P}_X = \{\frac{2}{5},\frac{1}{3},\frac{1}{5},%
\frac{1}{15}\}$, with Huffman merging we get expected code length $(1+\frac{9}{15}+\frac{4}{15} = 1.87 bits)$, while 
greedy division has expected code length $(1+1 = 2 bits)$.

% conference papers do not normally have an appendix

% use section* for acknowledgement
%\section*{Acknowledgment}
%
%The authors would like to thank...
%

% trigger a \newpage just before the given reference
% number - used to balance the columns on the last page
% adjust value as needed - may need to be readjusted if
% the document is modified later
\IEEEtriggeratref{8}
\newpage
% The "triggered" command can be changed if desired:
\IEEEtriggercmd{\enlargethispage{-5in}}

% references section

% can use a bibliography generated by BibTeX as a .bbl file
% BibTeX documentation can be easily obtained at:
% http://www.ctan.org/tex-archive/biblio/bibtex/contrib/doc/
% The IEEEtran BibTeX style support page is at:
% http://www.michaelshell.org/tex/ieeetran/bibtex/
%\bibliographystyle{IEEEtran}
% argument is your BibTeX string definitions and bibliography database(s)
%\bibliography{IEEEabrv,../bib/paper}
%
% <OR> manually copy in the resultant .bbl file
% set second argument of \begin to the number of references
% (used to reserve space for the reference number labels box)

% that's all folks
\end{document}